\documentclass[preprint,aps,notitlepage, nofootinbib,superscriptaddress]{revtex4}
\RequirePackage{xspace} 
\usepackage{booktabs} 
\usepackage{amsmath,amssymb}
\usepackage{mathrsfs}
\usepackage{bm}
\usepackage{algpseudocode,algorithm}

\newcommand{\fig}{Fig.}
\newcommand{\tab}{Tab.}

\newcommand{\e}{\text{e}} 
\newcommand{\tr}{\mathrm{tr}\,}  
\newcommand{\Tr}{\mathrm{Tr}\,}  
\newcommand{\av}[1]{\left\langle #1 \right\rangle} 
\renewcommand{\bar}[1]{\overline{#1}}%
\newcommand{\br}{\notag\\&\;\;\;\;\;\;}
\newcommand{\nconf}{ {N_\text{trj}} }
\usepackage[dvipdfmx]{color}
\RequirePackage{ifpdf}
\ifpdf
  \usepackage[pdftex]{graphicx}
\else
  \usepackage[dvipdfmx]{graphicx}
\fi
\usepackage{cases}
\usepackage[utf8]{inputenc}
\usepackage[T1]{fontenc}
\usepackage[english]{babel}
\begin{document}
%
%
\title{Self-learning Monte-Carlo for non-abelian gauge theory with dynamical fermions}
\preprint{\today}
\author{Yuki Nagai}
\email[]{nagai.yuki@jaea.go.jp}
\affiliation{Mathematical Science Team, RIKEN Center for Advanced Intelligence Project (AIP),1-4-1 Nihonbashi, Chuo-ku, Tokyo 103-0027, Japan}
\affiliation{CCSE, Japan  Atomic Energy Agency, 178-4-4, Wakashiba, Kashiwa, Chiba, 277-0871, Japan}
\author{Akinori Tanaka}
\email[]{akinori.tanaka@riken.jp}
\affiliation{interdisciplinary Theoretical \& Mathematical Sciences Program (iTHEMS) RIKEN 2-1, Hirosawa, Wako, Saitama 351-0198, Japan}
\affiliation{Mathematical Science Team, RIKEN Center for Advanced Intelligence Project (AIP),1-4-1 Nihonbashi, Chuo-ku, Tokyo 103-0027, Japan}
\affiliation{Department of Mathematics, Faculty of Science and Technology, Keio University, 3-14-1 Hiyoshi, Kouhoku-ku, Yokohama 223-8522, Japan}
\author{Akio Tomiya}
\email[]{akio.tomiya@riken.jp}
\affiliation{RIKEN/BNL Research center, Brookhaven National Laboratory, 
Upton, NY, 11973, USA}
\begin{abstract}
In this paper, we develop the self-learning Monte-Carlo (SLMC) algorithm for non-abelian gauge theory with dynamical fermions in four dimensions to resolve the autocorrelation problem in lattice QCD. We perform simulations with the dynamical staggered fermions and plaquette gauge action by both in HMC and SLMC for zero and finite temperature to examine the validity of SLMC. We confirm that SLMC can reduce autocorrelation time in non-abelian gauge theory
and reproduces results from HMC. For finite temperature runs, we confirm that SLMC reproduces correct results with HMC, including higher-order moments of the Polyakov loop and the chiral condensate. Besides, our finite temperature calculations indicate that four flavor QC${}_2$D with $\hat{m} = 0.5$ is likely in the crossover regime in the Colombia plot.
\end{abstract}
\pacs{}

\maketitle


\section{Introduction}
For more than 40 years, numerical calculations of lattice QCD have been an established technique to calculate QCD observables including non-perturbative effects \cite{Creutz:1980zw}.
Lattice QCD is based on the Markov chain Monte-Carlo (MCMC) algorithms with the detailed balance condition, which guarantees the MCMC process's convergence, and it is crucial for dealing with the dimensionality of the path integral for lattice QCD. 
From a practical point of view, guaranteeing the convergence is inevitable because it allows for uses of the computational result in precise phenomenology. 
(See \cite{Aoki:2019cca, Miura:2019xtd} and references therein, for example).

A MCMC algorithm for lattice QCD is required following three conditions.
The first condition is, convergence of the Markov chain update.
Typically, one requires the detailed balance condition because
it is a sufficient condition for the convergence.
The second condition is applicability for a non-abelian gauge system with dynamical fermions because QCD consists of gluons and quarks, and we must guarantee the gauge invariance.
The third condition is no-bias in calculations. For example, the hybrid algorithm or R algorithm can deal with
dynamical fermions, but it has a bias \cite{Kennedy:2006ax}, and it is not favored. Another example is the molecular dynamics,
which can generate configurations for lattice QCD by itself, but its ergodicity is not guaranteed.

{\it De facto} standard algorithms for gauge theories with dynamical fermions
are the hybrid Monte-Carlo (HMC) \cite{DUANE1987216}
and its variant rational hybrid Monte-Carlo (RHMC) \cite{Clark:2006wq}
because they satisfy the three conditions above.
The basic idea of (R)HMC is based on the Metropolis algorithm with fictitious 
hamiltonian dynamics, including Gaussian momentum, along with a fictitious time.
Generally speaking, a Metropolis algorithm enjoys high efficiency if the theory's energy function does not change so much during the update process.
(R)HMC uses the molecular dynamics with a reversible symplectic integrator, and it preserves the energy function mostly. 
The Gaussian momentum guarantees ergodicity of the algorithm. In addition to this update process allows us to include fermions through the pseudofermion trick.
Despite these advantages, (R)HMC suffers from unavoidable critical slowing down problem with several parameter regimes \cite{Schaefer:2010hu, Berg:1992qua, Ding:2020inp},
and thus to resolve the problem, a new algorithm is demanded.

On the other hand, recent progress of machine (ML) learning drives application it to lattice field theory \cite{Mehta_2019, Tanaka_2017, Bluecher:2020kxq, yoon2016estimation, Yoon_2019, Zhang_2020, nicoli2020estimation, tanaka2017reduction, Pawlowski_2020, Zhou_2019, Albergo_2019, Kanwar_2020, boyda2020sampling,
Alexandru_2017,Shanahan_2018,kades2019spectral,yoon2019machine,offler2019news,zhang2020pion,boyda2020machinelearning,gal2020baryons}.
There are three typical usages of machine learning in this context.
The first is making a detector of phase boundaries \cite{Tanaka_2017, Wetzel:2017ooo, Bluecher:2020kxq}.
The second is calculation for observable to reduce numerical cost \cite{yoon2016estimation, Yoon_2019, Zhang_2020, nicoli2020estimation}.
The third is configuration generation for field theory \cite{tanaka2017reduction, Pawlowski_2020, Zhou_2019, Albergo_2019, Kanwar_2020, boyda2020sampling}.
These applications are inspired by similarities between configuration generation and image generation but except for flow-based algorithm
\cite{Albergo_2019, Kanwar_2020, boyda2020sampling},
they do not enjoy the convergence theorem. 
This means that one has to compare observables one by one with observables generated by a ``legal algorithm'' like HMC.

Self-learning Monte-Carlo(SLMC) is a machine-learning 
based configuration generation algorithm \cite{Liu_2017,Nagai2017,Shen_2018,Chen2018,Nagai_2020,Liu2016,Nagai2020SLHMC}.
Originally it has been developed for a classical spin model \cite{Liu_2017}
but it works widely among quantum model in condensed matter physics and quantum chemistry \cite{Nagai2017,Nagai_2020}.
SLMC employs updates using a tunable action with an efficient update algorithm and the Metropolis -- Hastings test, and then, it becomes an exact algorithm.
One of the advantages of SLMC is interpretability; 
we can use our insight of the original theory to construct the tunable action.
Besides, one can also use neural networks
to evaluate the tunable action in the algorithm \cite{Shen_2018,Nagai_2020}
by giving up interpretability in order to achieve better acceptance.
We develop SLMC for non-abelian gauge field with dynamical
fermions.

From 90s, systems with dynamical fermions using hopping parameter expanded action have been investigated \cite{Irving_1997, Irving_1998, Duncan_1998, Duncan_1999a, Duncan_1999b, Hasenbusch_2018}.
They used truncated determinant to perform simulations for the Schwinger model and zero temperature QCD.
Here we clarify the difference between the present work and these studies.
First, we perform an extensive study for volume dependence and action dependence, and we find that the Polyakov loop in the effective action improves acceptance.
Second, we (re-)formulated the algorithm in the context of ML application.
It clarifies the meanings and possible extension of the algorithm.
Thirds, we confirm our algorithm reduce autocorrelation for a finite temperature system.
As an similar idea, the multi-boson algorithm \cite{Borrelli_1996, Alexandrou_2000} is known.
It uses the Metropolis algorithm with polynomially expanded fermion action.
But we do not use pseudofermion field to update the gauge field in the present work.
We leave that idea for future study.

In this work, we perform simulations with
two-color QCD with dynamical fermions in four dimensions (QC$_2$D)
using HMC and SLMC. 
Besides, we apply our algorithm to investigate QC$_2$D with 4 flavors
phase diagram associated the Polyakov loop in heavy mass regime
at finite temperature\footnote{
The one-loop beta function for SU($N_c=2$) gauge theory with $n_{f}=4$ fundamental matters
is 
$\beta(g)=-\frac{g^{3}}{(4 \pi)^{2}} \left(\frac{11}{3} N_c-\frac{2}{3} n_{f} \right)
=-\frac{g^{3}}{(4 \pi)^{2}} 4.6  < 0$,
thus, this theory is asymptotic free.
In addition, this theory is not infrared conformal \cite{Kaczmarek_1999}.
Thus, QC${}_2$D with $n_f=4$ has qualitatively same nature to conventional QCD.  
}.
In that regime, HMC is suffered from long autocorrelation problem of the Polyakov loop.
We find that, both in zero and finite temperature,
SLMC has smaller autocorrelation time than HMC
and gives consistent results with correct cumulants.

This paper is organized as follows. 
In section 2, we review self-learning Monte-Carlo from Metropolis -- Hasting algorithm.
In section 3, we explain our target system and our effective action in our calculations.
In section 4, we introduce our results at zero temperature runs.
In section 5, we show results for an application of SLMC to a finite temperature system.
In section 6, we summarize our results. 

\section{Self-learning Monte-Carlo}
Here we briefly derive the self-learning Monte-Carlo algorithm from the Metropolis--Hasting algorithm for gauge theories
to be self-contained.
\subsection{Metropolis-Hasting algorithm}
To generate configurations with a Markov-chain, we need to design a conditioned probability $P(B|A)$, where $A$ is a seed configuration
and $B$ is a generated configuration.
The Metropolis--Hasting algorithm is based on the detailed balance condition, 
\begin{align}
W_A P(B|A) ={W_B }P(A|B),
\end{align}
where $A$ and $B$ are a label for configurations
and $W_A$ is the Boltzmann weight proportional to $\e^{-S[U_A]}$ for a configuration $U_A$ with action $S \in \mathbb{R}$.
This guarantees configurations which are generated by the algorithm 
are distributed with the desired distribution associated with $W \propto  \e^{-S}$.
The Metropolis--Hasting algorithm is defined by following update rule,
\begin{align}
P_\text{MH}(B| A) \equiv \min\left(1, \frac{W_B T(A| B)}{W_A T(B| A)}  \right) T(B|A),
\end{align}
where $T(B|A)$ represents a proposing process,
which makes a candidate configuration $B$ from $A$ and
$\min(\cdots)$ denotes the Metropolis test.
Note that the process $T(B| A)$ does not have to be reversible in spite
of the molecular dynamics in HMC has to be reversible.
If one requires the reversibility to $T(B| A)$, it is reduced to the simplest Metropolis algorithm.
The detailed balance condition can be reproduced from the update algorithm:
\begin{align}
P_\text{MH}(B| A) 
&=  \frac{W_B T(A| B)}{W_A T(B| A)}  \min\left(1, \frac{W_A T(B|A)}{W_B T(A|B)} \right) T(B| A),\\
&=  \frac{W_B }{W_A } P_\text{MH}(A| B).
\end{align}
In the first line we used an identity $\min(1,c) = c \min(1,1/c)$ and $c$ is a non-negative real number.
In this manner, we can reproduce the detailed balance condition
from the Metropolis-Hasting update rule. 

\subsection{Self-learning Monte-Carlo}
Self-learning Monte Carlo (SLMC) is one of the Metropolis-Hasting algorithms.
It is obtained by regarding $T(B| A) = P^\text{eff}_\theta(B| A)$, which is an update algorithm associated with an effective action with a set of parameters $\theta$ like couplings in the effective model. 
The parameter set  $\theta$ is determined by a machine learning estimator, and we use the linear regression in this work.
In the framework of SLMC, the effective action has to have an efficient update algorithm with the detailed balance for the effective action,
\begin{align}
\frac{W_A^\text{eff,$\theta$} }{ W_B^\text{eff,$\theta$} }  
= \frac{ P^\text{eff}_\theta(A| B)  }{ P^\text{eff}_\theta(B| A)  } \label{eq:detailed_for_eff}.
\end{align}
Let us regard $T(B| A)$ as $P^\text{eff}_\theta(B| A)$ in the Metropolis Hasting algorithm,
\begin{align}
P_\text{SLMC}(B| A)
&\equiv \min\left(1, \frac{W_B P^\text{eff}_\theta(A| B)}{W_A P^\text{eff}_\theta(B| A) }  \right) P^\text{eff}_\theta(B| A) ,
\end{align}
and this is the update rule for the self-learning Monte-Carlo.
The detailed balance can be derived from this update,
\begin{align}
P_\text{SLMC}(B|A) 
&= \frac{W_B P^\text{eff}_\theta(A| B)}{W_A P^\text{eff}_\theta(B| A) }  \min\left(1, \frac{W_A P^\text{eff}_\theta(B| A) }{W_B P^\text{eff}_\theta(A| B)}  \right) P^\text{eff}_\theta(B| A) ,\\
&= \frac{W_B }{W_A }  \min\left(1, \frac{W_A P^\text{eff}_\theta(B| A) }{W_B P^\text{eff}_\theta(A| B)}  \right) P^\text{eff}_\theta(A| B) ,\\
&= \frac{W_B }{W_A }  P_\text{SLMC}(A| B).
\end{align}
The self-learning update satisfies the detailed balance and it will converge.
The modified Metropolis test can be evaluated with the effective action because of \eqref{eq:detailed_for_eff}
and,
\begin{align}
\min\left(1, \frac{W_B P^\text{eff}_\theta(A| B)}{W_A P^\text{eff}_\theta(B| A ) }  \right)
&=
\min\left(1, \frac{W_B }{W_A  } \frac{ W_A^\text{eff,$\theta$} }{W_B^\text{eff,$\theta$} } \right) 
=
\min\left(1, \frac{W_B/W_B^\text{eff,$\theta$} }{W_A/W_A^\text{eff,$\theta$} } \right).
\end{align}
Here we do not have to assume reversibility for $T=P^\text{eff}_\theta$ again
but it requires the detailed balance for the effective action instead.
Note that, if one uses an effective action which differs from
the target system's one, the acceptance rate could be low
but even such case the expectation values still converge into the exact values
with long autocorrelation ({\it i.e.}, large statistical error)
because this is an exact algorithm.

\section{Numerical setup}
In this section, we introduce our numerical setup.
Though out this paper, we show every quantity in lattice unit.
We perform simulations with $SU(2)$ plaquette action with 4 tastes standard staggered fermions 
for implementational simplicity. 
Our target system is described by the action $S[U,\bar{\chi},\chi] $,
\begin{align}
S[U,\bar{\chi},\chi] 
&= S_g[U] +\sum_n (\bar\chi M[U] \chi)_n, \label{eq:full_action_in_fermion}
\end{align}
where $S_g[U]$ is the Wilson plaquette action,
\begin{align}
S_g[U] &= \beta \sum_n 
\sum_{\mu=1}^4 \sum_{\nu>\mu} \left( 1 - \frac{1}{2}  \tr U_{\mu\nu}(n) \right) ,\\
U_{\mu\nu}(n) &= U_\mu(n) U_\nu(n+\hat\mu) U_\mu^\dagger(n+\hat\nu) U_\nu^\dagger(n),
\end{align}
and $M[U]$ is a massive staggered Dirac operator,
\begin{align}
M[U]\chi &= \frac{1}{2}\sum_{\mu=1}^4 \eta_{\mu} (n) \left[
U_\mu(n)\chi(n+\hat\mu) - U_\mu^\dagger(n-\hat\mu)\chi(n-\hat\mu)
\right] + \hat{m}\chi(n),
\end{align}
and $\eta_{\mu} (n)$ is the staggered factor.
$U_\mu(n)$ is a link variable for $SU(2)$ gauge field,
and $\chi(n)$ is a single component spinor field.
$\hat{m}$ is a quark mass in the lattice unit.

All numerical calculations for HMC and SLMC are performed by Julia \cite{bezanson2017julia}
and the code is developed by ourselves.
In addition, we implement automatic generation of heatbath code which
generates staples from given loop operators \cite{LQCDJulia}.
This enables us to investigate effective action with various types of extended loops.
However, in present work, we employ the effective action \eqref{eq:effective_action}
for simplicity and study with complicated loops leave it as a future work.

\subsection{SLMC setup}
SLMC update probability\footnote{
For comparison, HMC can be written as,
\begin{align*}
P_\text{HMC}(U'| U) = \min\left(1, \frac{e^{-(S[U',M[U']\phi] + \frac{1}{2}{\pi'}^2  ) }  }
{e^{-(S[U,M[U]\phi] + \frac{1}{2}\pi^2 )} } \right) T^{\theta_\text{MD}}_\text{MD}(U',\pi',\phi|U,\pi,\phi)\delta(\phi-M[U]\eta)P_G(\eta)P_G(\pi),
\end{align*}
where $\theta_\text{MD}$ is a set of parameters for the molecular dynamics
{\it i.e.}, the step size $\epsilon_\text{MD-step}$, choice of MD evolution time, and multi-scale integration splitting.
$P_G(\cdot)$ is the Gaussian distribution.
$T^{\theta_\text{MD}}_\text{MD}(U',\pi',\phi|U,\pi,\phi)$ must be reversible under fictitious time, 
which is realized by a symplectic integrator.
}
for the lattice gauge theory is written as,
\begin{align}
P_\text{SLMC}(U'|U) = \min\left(1, \frac{e^{-(S[U'] - S^{\theta}_\text{eff}[U']) }  }
{e^{-(S[U] - S^{\theta}_\text{eff}[U])} } \right) P^{\theta}_\text{eff}(U'|U) ,
\end{align}
where our target action in bosonic language is given by, 
\begin{align}
S[U] = S_g[U] + S_f[U],
\end{align}
where $S_g[U]$ is the plaquette gauge action and,
\begin{align}
S_f[U] = -\log \det M^\dagger M,
\end{align}
and $ S^{\theta}_\text{eff}[U]$ is an effective action with tunable parameters,
which is only consisted by loop operators {\it e.g.}, plaquette and rectangular and so on.
We use exact diagonalization to evaluate the Dirac operator in this paper for simplicity
and this can be improved by using stochastic estimator in the reweighting technique \cite{Hasenfratz_2008, DeGrand_2008}.
We call this algorithm as self-learning Monte-Carlo (SLMC) hereafter.

In practice, we employ heavy mass expanded fermion action with truncation \cite{Saito_2011} for updates in SLMC.
Our effective action is consisted by the plaquette, rectangular loops and the Polyakov loop for $\mu=1,2,3,4$ directions.
Let us denote $\vec{n}=(n_1,n_2,n_3)$ is spatial coordinate and $n_4$ is the temporal direction
and let lattice size be $N_1,N_2,N_3$ and $N_4$ for spatial and temporal extent, respectively.
The effective action is,
\begin{align}
S^{\theta}_\text{eff}[U] 
&=   \sum_n 
\left[
\beta_\text{plaq} \sum_{\mu=1}^4 \sum_{\nu>\mu} \left( 1 - \frac{1}{2}  \tr U_{\mu\nu}(n) \right)
+
\beta_\text{rect} \sum_{\mu=1}^4 \sum_{\nu\neq\mu} \left( 1 - \frac{1}{2}  \tr R_{\mu\nu}(n) \right)
\right]\br
+\beta^{\mu=1}_\text{Pol}  \sum_{n_2,n_3,n_4}  \tr \big[\prod_{n_1=0}^{N_1-1} U_1(\vec{n},n_4)  \big]
+\beta^{\mu=2}_\text{Pol}  \sum_{n_1,n_3,n_4}  \tr \big[\prod_{n_2=0}^{N_2-1} U_2(\vec{n},n_4)  \big]\br
+\beta^{\mu=3}_\text{Pol}  \sum_{n_1,n_2,n_4}  \tr \big[\prod_{n_3=0}^{N_3-1} U_3(\vec{n},n_4)  \big]
+\beta^{\mu=4}_\text{Pol}  \sum_{n_1,n_2,n_3}  \tr \big[\prod_{n_4=0}^{N_4-1} U_4(\vec{n},n_4)  \big],
\label{eq:effective_action}
\end{align}
where $n=(\vec{n},n_4)$ and $\vec{n} = (n_1,n_2,n_3)$
and $R_{\mu\nu}(n)$ is a rectangular Wilson loop,
\begin{align}
R_{\mu\nu}(n) &= U_\mu(n)U_\mu(n+\hat\mu) U_\nu(n+2\hat\mu)  U_\mu^\dagger(n+\hat\mu+\hat\nu) U_\mu^\dagger(n+\hat\nu) U_\nu^\dagger(n),
\end{align}
and $\theta = \{\beta_\text{plaq}, \beta_\text{rect}, \beta^{\mu=1}_\text{Pol}, \beta^{\mu=2}_\text{Pol}, \beta^{\mu=3}_\text{Pol}, \beta^{\mu=4}_\text{Pol}\}$
are determined by a linear regression with prior HMC run\footnote{
Prior HMC runs are not mandatory. One can improve the parameters within the SLMC runs.
Most of our simulations in this paper are done in this way.
If the effective action is far from the target system, a prior HMC run is necessary to avoid inefficiency. 
}.
In general cases, one can include more and more extended loops to improve the acceptance rate.
This algorithm is based on ML but has interpretability. Namely, coefficients in the effective action have physical meaning.
Note that this effective action for SLMC does not require systematic expansion so that we can drop several terms. 
This operation does not bring any bias but causes inefficiency of simulation.

We choose $P^{\theta}_\text{eff}(U'|U)$ as the heatbath algorithm with whole extended loops in $S^{\theta}_\text{eff}[U] $.
Our strategy in the present work to overcome the critical slowing down is to increase the number of heatbath updates or overrelaxation in the SLMC update process.
The critical slowing down depends on both of the update algorithm and the criticality of the system, and criticality is unavoidable.
We overcome the critical slowing down by somewhat brute force way; repeating cheap updates\footnote{
This strategy is the same spirit of the all mode averaging (AMA) \cite{Blum_2013}.
AMA reduces statistical error using many ``sloppy'' (cheap) calculation,
and bias from sloppiness corrected by 
a few costly bias correction term.
}.

The acceptance rate can be estimated {\it a priori} by a loss of the regression and,
\begin{align}
\text{Acceptance rate} \sim \exp( - \sqrt{L_2})
\end{align}
and $L_2$ is a loss of the regression for the effective action \cite{Shen_2018}, 
which is similar to Karsch formula \cite{Gupta:1990ka, Irving_1997}.
Namely, the acceptance rate can be reduced controlled by adding more and more extended loops
as improving the linear regression.

Here we note how we can compare different algorithms.
Generally, it is not fair to compare two different algorithms in terms of the elapsed time since the elapsed time depends on the architecture of machines and technical details of implementations. 
In this paper, we count the number of most costly expensive parts in each algorithm.
In HMC case, the number of inversion of the Dirac operator, which occurs each molecular dynamics step and the Metropolis step is the expensive part.
The conjugation gradient (CG) method is usually used to calculate the inversion of the Dirac operator, which has many matrix-vector operations. 
On the other hand, in our SLMC case, we count the number of the Metropolis test,
which contains determinant calculation for the action of fermions.
As we mentioned before, this can be replaced by a stochastic estimator
and it reduces numerical cost, but it still has the highest cost.
A stochastic estimator includes the calculation of the inversion of the Dirac operator, which is similar to the CG method. 
In this work, however, we count the number of Metropolis test both in HMC and SLMC for simplicity. This comparison is not fair for SLMC, but still, it gives better results than HMC in terms of the autocorrelation.

\subsection{Observables}
In the present work, we measure plaquette, rectangular Wilson loop, and the Polyakov loop to check the consistency of the algorithm.
In our finite temperature application,  we calculate a susceptibility (second-order cumulant)  and the Binder cumulant, which is a fourth order moment, for the plaquette, rectangular Wilson loop, Polyakov loop,
and chiral condensates as a function of $\beta$ to check the consistency of the algorithm for possible biases in higher moments in addition to the mean values. 

The Polyakov loop along with the temporal direction 
is a good indicator of the confinement-deconfinement transition,
\begin{equation}
\av{L} = \frac{1}{N_\sigma^3 }\left\langle \sum_{\vec{n}}\mathrm{Tr}\prod_{n_4}U_4(n_4,\vec{n})\right\rangle,
\end{equation}
where $N_\sigma$ is the spatial size of the lattice.
The Polyakov loop susceptibility is,
\begin{align}
\chi_{L} = \av{L^2} - \av{L}^2.
\end{align}
We also analyze the Binder cumulant $B^4_L$ \cite{binder1981critical},
which is defined by,
\begin{align}
B^4_{L} (\beta) = \frac{\langle (\delta L)^4 \rangle} {\langle(\delta L)^2 \rangle^2},
\end{align}
where $\delta L=L-\langle L \rangle$.
Binder cumulant is an indicator of the order of phase transition.
If it takes $B^4 = 3.0$, that point does not have any singularity (crossover).
If it takes $1<B^4 < 3.0$, that point is the second-order phase transition, and the value is related to the universality class.
If it takes $B^4 = 1.0$, that point is the first-order phase transition.
In addition to the Polyakov loop,
we calculate four tastes chiral condensate,
\begin{align}
 \av{\bar{\psi}\psi} 
= \frac{1}{N_\sigma^3 N_\tau}
\av{ \Tr\frac{1}{D+m}  } ,
\end{align}
where $\Tr$ indicates trace over all index in the Dirac operator,
and its higher order moments as well as for the Polyakov loop.

Autocorrelation time is a measure of correlations between configurations,
which quantifies the inefficiency of an MCMC algorithm.
The decay of the autocorrelation function gives autocorrelation time, but
the autocorrelation function itself is a statistical object,
so we cannot determine the autocorrelation exactly. 
Instead, we calculate the approximated autocorrelation function
\cite{Wolff:2003sm, Madras:1988ei} defined by,
\begin{align}
\bar\Gamma(\tau) = \frac{1}{\nconf - \tau} \sum_{c}^\nconf (O_c-\bar{O})(O_{c+\tau}-\bar{O}),
\label{autocorrelation_function}
\end{align}
where $O_c = O[ U^{(c)} ]$ is the value of operator $O$ for the $c$-th configuration $U^{(c)}$ and $\tau$ is the fictitious time of HMC.
$\nconf$ is the number of trajectories.
Conventionally, the normalized  autocorrelation function $\bar\rho(\tau)=\bar\Gamma(\tau) /\bar\Gamma(0) $
is used.

The integrated autocorrelation time $\tau_\text{int}$ approximately quantifies 
effects of autocorrelation. This is given by,
\begin{align}
\tau_\text{int} = \frac{1}{2} + \sum^{W}_{\tau=1} \bar\rho(\tau).
\label{tau_int}
\end{align}
We regard two configurations separated by $2\tau_\text{int} $ as independent ones.
In practice, we determine a window size $W$ as a first point $W = \tau $ where $\bar{\Gamma}(\tau)<0$ for the smallest $\tau$.
The statistical error of integrated autocorrelation time
is estimated by the Madras--Sokal formula \cite{Madras:1988ei, Luscher:2005rx},
\begin{align}
\av{\delta \tau_\text{int}^2 }
\simeq
\frac{4W + 2}{\nconf} \tau_\text{int}^2, \label{eq:ac_time_error}
\end{align}
We use the square root of \eqref{eq:ac_time_error} to estimate the error on the autocorrelation time.
It is obvious that the autocorrelation is observable dependent,
and we focus on the autocorrelation from the Polyakov loop since
we perform simulations with  lattice with rather coarse lattices\footnote{
We checked the topological charge and its autocorrelation, but that is not relevant for our lattice spacing and system size.
}.

\section{Algorithm analysis}
We compare results from SLMC and HMC with a base-line parameter set:
$N_\sigma^3 \times N_\tau=6^3\times6$, $\hat{m}=0.5$, $\beta=2.5$.
Our effective action contains plaquette, rectangular, Polyakov loops for every direction
as we explained \eqref{eq:effective_action}.
We discard $O(100)$ trajectory from the analysis for thermalization,
and the number of analyzed configurations\footnote{
All of our measurements are done on the fly and performed every trajectory.
} is $O(1000)$-$O(10000)$.
We estimate statistical error using the Jackknife method.
The bin size of the Jackknife method is taken to be the number of Jackknife samples 10 in the histogram, and others are taken to be larger than the autocorrelation time for each observable.

\begin{figure}[h]
\begin{center}
\begin{minipage}{0.45\hsize}
\includegraphics[scale=0.45]{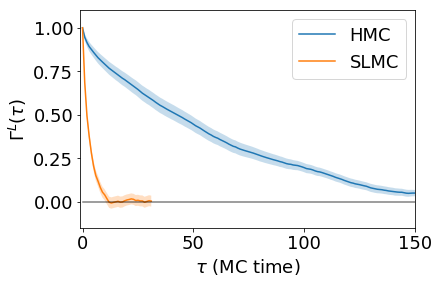}
\end{minipage}
\end{center}
  \caption{
  Autocorrelation function for the Polyakov loop
  for $\beta = 2.5$, $N_\sigma^3\times N_\tau = 6^4$, $\hat{m}= 0.5$.
  We measure the Monte-Carlo time $\tau$ as a unit of the Metropolis test.
  The bands indicate one sigma error bar. Please see main text for meaning of
  the horizontal axis.
  \label{fig:acfuncp_L6_m05_b25_base}}
\end{figure}

Before detail comparison,
we compare HMC and SLMC
for $\beta = 2.5$, $N_\sigma^3\times N_\tau = 6^4$, $\hat{m}= 0.5$
(\fig \ref{fig:acfuncp_L6_m05_b25_base}).
The horizontal axis (MC time) is counted as the number of the Metropolis test as we explained above.
The integrated autocorrelation time for HMC and SLMC are $\tau^\text{HMC} = 62(24)$ and $\tau^\text{SLMC} = 4.0(4)$, respectively.
We choose the number of overrelaxation as 10 and the number of heatbath as 100 in this case.

\begin{figure}[h]
\begin{center}
\begin{minipage}{1.0\hsize}
\includegraphics[scale=0.33]{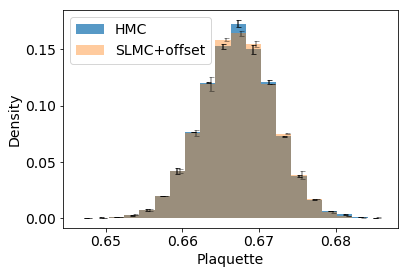}
\includegraphics[scale=0.33]{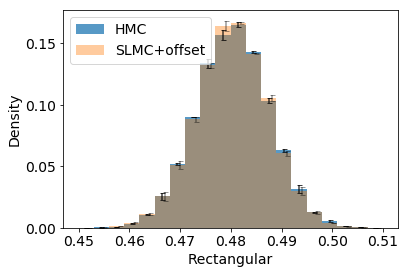}
\includegraphics[scale=0.33]{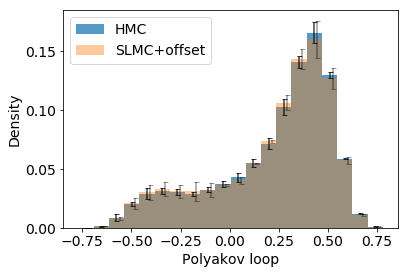}
\end{minipage}
\end{center}
  \caption{
Histogram of results from HMC and SLMC for $\beta = 2.5$, $N_\sigma^3\times N_\tau = 6^4$, $\hat{m}= 0.5$.
To distinguish these results, we shift results for SLMC to the right. Error bar is estimated by the Jackknife method.
({\it Left}) Plaquette.
({\it Middle}) Rectangular Wilson loop.
({\it Right}) Polyakov loop.
\label{fig:t0_baseline}}
\end{figure}

We show results from HMC and SLMC for
plaquette, rectangular Wilson loop, and Polyakov loop in 
histogram with statistical error in \fig \ref{fig:t0_baseline}.
One can see that all quantities from SLMC are consistent with ones from HMC.
The shape of the histogram for the Polyakov loop indicates the effects of dynamical fermions because if it is quenched, the Polyakov loop is symmetric under $\mathbb{Z}_2$ reflection if the statistics are large enough.

Other quantities and setups
are summarized the results in \tab\ref{tab:zero_temp}. 
ID 0--7 in the table are results from HMC and various SLMC with
$N_\sigma^3\times N_\tau = 6^4$, $\beta = 2.5$, and $\hat{m} = 0.5$.
SLMC\_nup100 means SLMC with 100 times heatbath update.
SLMC\_all means that SLMC with effective action, including $3\times1$-rectangular, chair, and crown operators.
SLMCnor01 and SLMCnor20 mean SLMC with 1 and 20 time overrelaxation after the heatbath update, respectively.
SLMCplq means SLMC with an effective action, which includes only plaquette term.
SLMCplqrct means SLMC with an effective action, which includes plaquette and the rectangular term.
All of the results from various SLMC are consistent with ones from HMC, as expected.
Besides, SLMCplq contains only one term, but it achieves roughly 60 \% acceptance
and with consistent results. 
Comparing to the acceptance rate for SLMCplqrct and SLMC\_all,
improvement of effective action, namely adding more loop operators, 
the data shows that adding loops improves the acceptance rate.

Beta dependence are summarized in ID 8--13 in \tab\ref{tab:zero_temp}. 
We vary $\beta$ as $0.8, 1.2, 4.0$ both in HMC and SLMC.
The acceptance rate in SLMC for $\beta=0.8$ is slightly low, but it gives consistent results.
$\beta$ dependence is correctly reproduced.

Results from lighter mass $\hat{m} = 0.05$ are summarized in ID 14--16 in \tab\ref{tab:zero_temp}. 
In this case, for SLMC, acceptance is low (30-40\%), but it still gives consistent results to the ones in HMC.
The acceptance rate for SLMC\_polys  (effective action with char, crown, and $3\times1$ Wilson loop) is improved
from SLMC.
ID 17--18 in the table are results for $\hat{m} = 0.1$.
The tendency of acceptance rate is the same to $\hat{m}=0.05$ but slightly better as expected.

We examine volume dependence in the table (ID 19--21) and  the volume is taken to $N_\sigma^3\times N_\tau = 8^4$.
Acceptance rate is lower than the SLMC with $N_\sigma^3\times N_\tau = 6^4$.
This is similar to what happens for the reweighting, but in our case, thanks to the tunable parameters,
inefficiency is not drastic.
Examinations for smaller volume $N_\sigma^3\times N_\tau = 4^4$ are summarized in ID 22--23.
The acceptance rate reaches to 80\% for SLMC, and this is also expected.

In summary, SLMC can reproduce results from HMC even with plaquette effective action.
The number of terms in the effective action affects to acceptance rate.
For larger volumes and small mass, the acceptance rate becomes small. 
This can be improved by adding more and more terms to the effective action.

\begin{table}[htb]
\begin{tabular}{cccccc||cc|ccc}
\hline %
ID &           ALG &  $N_\sigma$  &  $N_\tau$ &  $\beta$ &    $m$ & Acceptance & $N_\text{trj}$&       $\av{P}$ &        $ \av{R}$ &      $\av{L}$ \\
\hline\hline%
0  &          HMC &   6 &   6 &   2.5 &  0.50 &        0.65 &  50000 &   0.66718(5) &  0.48037(9) &    0.23(1) \\
1  &  SLMC\_nup100 &   6 &   6 &   2.5 &  0.50 &        0.72 &  48850 &   0.66711(1) &  0.48021(3) &   0.197(3) \\
2  &         SLMC &   6 &   6 &   2.5 &  0.50 &        0.73 &  50000 &   0.66718(3) &  0.48031(5) &    0.22(1) \\
3  &     SLMC\_all &   6 &   6 &   2.5 &  0.50 &        0.77 &  50000 &   0.66719(3) &  0.48034(5) &    0.19(1) \\
4  &    SLMCnor01 &   6 &   6 &   2.5 &  0.50 &        0.74 &  50000 &   0.66717(3) &  0.48032(5) &    0.23(2) \\
5  &    SLMCnor20 &   6 &   6 &   2.5 &  0.50 &        0.73 &  50000 &   0.66731(3) &  0.48054(4) &   0.191(9) \\
6  &      SLMCplq &   6 &   6 &   2.5 &  0.50 &        0.57 &  50000 &   0.66746(4) &  0.48074(5) &    0.18(1) \\
7  &   SLMCplqrct &   6 &   6 &   2.5 &  0.50 &        0.69 &  50000 &   0.66727(5) &  0.48046(7) &   0.211(8) \\
\hline%
8  &          HMC &   6 &   6 &   0.8 &  0.50 &        0.85 &  50000 &   0.20764(4) &  0.04444(3) &  0.0025(3) \\
9  &          HMC &   6 &   6 &   1.2 &  0.50 &        0.84 &  50000 &   0.30356(4) &  0.09382(3) &  0.0034(2) \\
10 &          HMC &   6 &   6 &   4.0 &  0.50 &        0.73 &  50000 &   0.80346(3) &  0.68011(4) &    0.65(1) \\
11 &         SLMC &   6 &   6 &   0.8 &  0.50 &        0.58 &  50000 &   0.20781(5) &  0.04454(5) &  0.0032(4) \\
12 &         SLMC &   6 &   6 &   1.2 &  0.50 &        0.59 &  50000 &   0.30365(4) &  0.09387(3) &  0.0033(7) \\
13 &         SLMC &   6 &   6 &   4.0 &  0.50 &        0.87 &  50000 &    0.8035(2) &  0.68016(3) &    0.58(4) \\
\hline%
14 &          HMC &   6 &   6 &   2.5 &  0.05 &        0.82 &  50000 &   0.67774(4) &  0.49772(5) &   0.437(6) \\
15 &         SLMC &   6 &   6 &   2.5 &  0.05 &        0.34 &  50000 &   0.67813(8) &   0.4982(1) &   0.436(7) \\
16 &   SLMC\_polys &   6 &   6 &   2.5 &  0.05 &        0.43 &  36350 &   0.67793(4) &  0.49798(5) &   0.446(7) \\
\hline%
17 &          HMC &   6 &   6 &   2.5 &  0.10 &        0.73 &  50000 &    0.6771(4) &  0.49666(5) &   0.428(6) \\
18 &         SLMC &   6 &   6 &   2.5 &  0.10 &        0.37 &  50000 &   0.67749(7) &  0.49732(9) &   0.438(5) \\
\hline%
19 &          HMC &   8 &   8 &   2.5 &  0.50 &        0.77 &  50000 &   0.66659(2) &  0.47916(2) &    0.01(1) \\
20 &         SLMC &   8 &   8 &   2.5 &  0.50 &        0.54 &   6630 &  0.66682(10) &   0.4795(2) &   -0.03(4) \\
21 &     SLMC\_all &   8 &   8 &   2.5 &  0.50 &        0.62 &   5300 &   0.66678(9) &   0.4794(1) &    0.04(3) \\
\hline%
22 &          HMC &   4 &   4 &   2.5 &  0.50 &        0.66 &  50000 &    0.6706(1) &   0.4878(2) &    0.64(2) \\
23 &         SLMC &   4 &   4 &   2.5 &  0.50 &        0.84 &  50000 &   0.67073(7) &  0.48792(9) &   0.656(7) \\
\hline %
\end{tabular}
\caption{
Simulation results for zero temperature with HMC and SLMC.
ALG indicates different algorithms, and please see the main text for details.
$\hat{m}$ indicates dimensionless quark mass.
Acceptance means the acceptance rate.
$N_\text{trj}$ is the number of trajectories except for the thermalization.
$\av{P}$, $\av{R}$, and $\av{L}$ mean expectation value of plaquette, rectangular Wilson loop,
and the Polyakov loop, respectively.
\label{tab:zero_temp}
}
\end{table}

\section{Application to finite temperature}
\subsection{Simulation setup}
Here we present an application of SLMC algorithm to a finite temperature system.
We perform simulation for  four flavor QC$_2$D with $\hat{m}=0.5$
in $N_\sigma^3\times N_\tau = 8^3\times 4$ lattice.
As we will mention later, quarks are not decoupled from the theory.
Our $\beta$ range is $\beta = 1$ -- $2.4$,
which contains a transition (crossover) point $\beta_c \sim 2.1$. 
We employ the Wilson plaquette gauge action and
the standard staggered fermion. 
In SLMC update, 20 times heatbath updates are used 
except for $\beta = 2.1$ while 100 times heatbath for $\beta = 2.1$.
The number of trajectory for HMC are 1000 for $\beta \neq 2.1$
and 20000 for $\beta = 2.1$ to see behavior of the Binder cumulant in detail.
The number of trajectory for SLMC are O(5000)--O(40000)
and please find details in \tab\ref{tab:fin_temp}.

At the heavy quark mass regime, the system expected to show
confinement/deconfinement transition (or crossover) associated with the Polyakov loop.
The Polyakov loop operator, along with the imaginary time direction, has 
the center symmetry, but it is broken at high temperatures.
It means that the system around or above the (pseudo-)critical temperature,
the system could be affected by long autocorrelation for the Polyakov loop.
We attempt to solve this autocorrelation by employing SLMC.
Meanwhile, we show that SLMC with effective action can 
treat detail of phase transition without bias.

\subsection{Simulation results}
\begin{figure}[h]
\begin{center}
\begin{minipage}{0.45\hsize}
\includegraphics[scale=0.45]{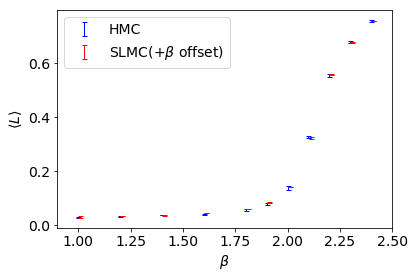}
\end{minipage}
\begin{minipage}{0.45\hsize}
\includegraphics[scale=0.45]{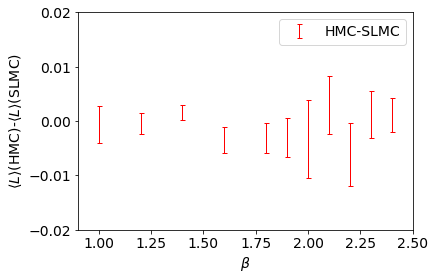}
\end{minipage}
\end{center}
  \caption{Polyakov loop as a function of $\beta$.
  To avoid overlapping of symbols, we slightly shift in $\beta$ for SLMC to the right in plots.
  ({\it Left}) Comparison plot for HMC and SLMC.
  ({\it Right}) Difference between Polyakov loop by HMC and SLMC.
  \label{fig:fintemp_polyakov_loop_ave_beta}}
\end{figure}

Here we show results at finite temperature.
At the heavy mass regime, Polyakov loop is a central observable.
Polyakov loop as a function of $\beta$ is shown in left panel of 
\fig \ref{fig:fintemp_polyakov_loop_ave_beta}
One can see that a transition point is around $\beta \sim 2.1$.
Right panel of 
\fig \ref{fig:fintemp_polyakov_loop_ave_beta}
is difference of results for the Polyakov loop from HMC and SLMC
and the values are consistent with 0.
Central values for other gluonic observables 
can be found in \tab\ref{tab:fin_temp}.

\begin{figure}[h]
\begin{center}
\begin{minipage}{0.45\hsize}
\includegraphics[scale=0.45]{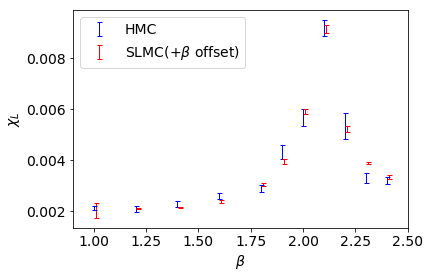}
\end{minipage}
\begin{minipage}{0.45\hsize}
\includegraphics[scale=0.45]{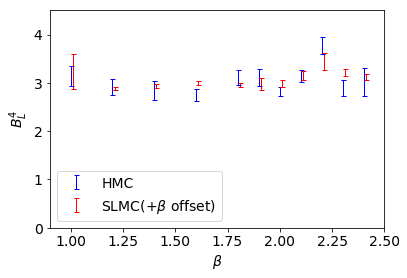}
\end{minipage}
\end{center}
  \caption{Susceptibility and the Binder cumulant for Polyakov loop as a function of $\beta$.
    To avoid overlapping symbols, we shift $\beta$ for SLMC in plots.
    ({\it Left}) Polyakov loop susceptibility. It has a peak $\beta \sim 2.1$.
  ({\it Right}) The Binder cumulant. It takes values about $ 3$ for all $\beta$ regime.
  \label{fig:fintemp_polyakov_loop_sus_beta}}
\end{figure}

Next, we show results for the susceptibility for Polyakov loop (left panel of \fig \ref{fig:fintemp_polyakov_loop_sus_beta}).
One can find a peak of around $\beta \sim 2.1$.
Besides, results from HMC and SLMC are consistent with each other.
Right panel of \fig \ref{fig:fintemp_polyakov_loop_sus_beta} is the Binder cumulant for Polyakov loop.
Results from HMC and SLMC are consistent with each other.
Besides, the Binder cumulant suggest that our quark mass $\hat{m}=0.5$ is not in the quenched regime
because in the quenched case, it takes
a value for the second-order phase
transition $B^4_L \approx 1.6$ \cite{Engels:1989fz,denbleyker2009finite}.

\begin{figure}[h]
\begin{center}
\begin{minipage}{0.45\hsize}
\includegraphics[scale=0.45]{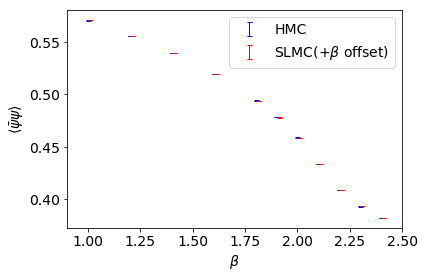}
\end{minipage}
\begin{minipage}{0.45\hsize}
\includegraphics[scale=0.45]{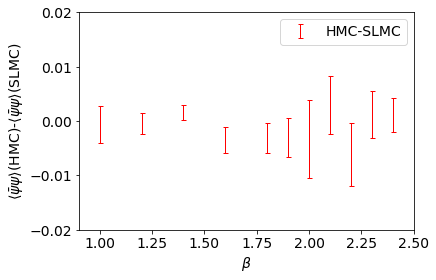}
\end{minipage}
\end{center}
  \caption{
Same plots of \fig \ref{fig:fintemp_polyakov_loop_ave_beta}
but for the chiral condensate.
  \label{fig:fintemp_chiral_ave_beta}}
\end{figure}

Next, we show results for the chiral condensates as a function of $\beta$ (Left panel of  \fig \ref{fig:fintemp_chiral_ave_beta}).
The transition point is located $\beta \sim 2.1$.
Right panel of \fig \ref{fig:fintemp_chiral_ave_beta}
is difference of results for the chiral condensates from HMC and SLMC.
Results show that central values are consistent with each other.

\begin{figure}[h]
\begin{center}
\begin{minipage}{0.45\hsize}
\includegraphics[scale=0.45]{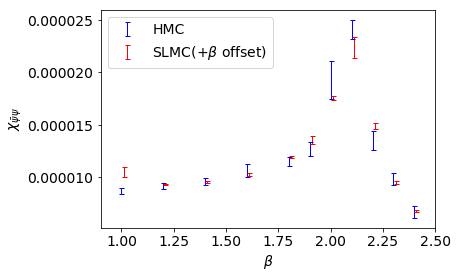}
\end{minipage}
\begin{minipage}{0.45\hsize}
\includegraphics[scale=0.45]{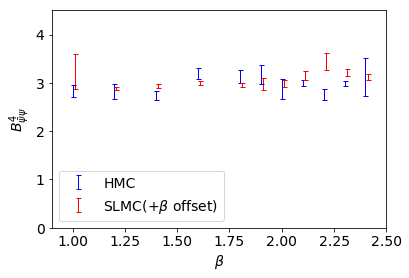}
\end{minipage}
\end{center}
  \caption{
Same plots of \fig \ref{fig:fintemp_polyakov_loop_sus_beta}
but for the chiral condensate.
  \label{fig:fintemp_chiral_sus_beta}}
\end{figure}

Results of higher cumulants can be found in \fig \ref{fig:fintemp_chiral_sus_beta}.
Left panel of \fig \ref{fig:fintemp_chiral_sus_beta} shows the susceptibility for the chiral condensates
while the right panel is Binder cumulant for the chiral condensates.
Results from HMC and SLMC are consistent with each other.

Our results of the Binder cumulant for the Polyakov loop 
and chiral condensates
(\fig \ref{fig:fintemp_polyakov_loop_sus_beta}
and \fig \ref{fig:fintemp_chiral_sus_beta}) at $\hat{m}=0.5$
do not show any indication of the second order (with three-dimensional Ising universality class\footnote{
Svetitsky and Yaffe have conjectured the confinement/deconfinement transition for pure SU(2) in 3+1 dimension
is the second order phase transition with 3D-Ising universality class \cite{Svetitsky:1982gs}
while the  chiral phase transition is first order for $n_f=4$ light quarks by Pisarski and Wilczek \cite{Pisarski:1983ms, Kaczmarek_1999}.
}).
This can be confirmed by performing a systematic study for large volumes, a finer scan of beta, and increasing statistics, but we leave this issue for a future study. 

\begin{table}[htb]
\begin{tabular}{cccccc||cc|ccc}
\hline %
ID &           ALG &  $N_\sigma$  &  $N_\tau$ &  $\beta$ &    $m$ & Acceptance & $N_\text{trj}$&       $\av{P}$ &        $ \av{R}$ &      $\av{L}$ \\
\hline\hline%
0  &             HMC &   8 &   4 &   1.0 &  0.5 &        0.88 &   1000 &    0.256(2) &    0.067(1) &   0.028(2) \\
1  &             HMC &   8 &   4 &   1.2 &  0.5 &        0.89 &   1000 &   0.3037(2) &   0.0939(1) &   0.031(2) \\
2  &             HMC &   8 &   4 &   1.4 &  0.5 &        0.87 &   1000 &    0.353(2) &   0.1269(2) &   0.036(1) \\
3  &             HMC &   8 &   4 &   1.6 &  0.5 &        0.87 &   1000 &   0.4048(3) &   0.1674(3) &   0.039(2) \\
4  &             HMC &   8 &   4 &   1.8 &  0.5 &        0.85 &   1000 &   0.4625(3) &   0.2204(3) &   0.055(3) \\
5  &             HMC &   8 &   4 &   1.9 &  0.5 &        0.85 &   1000 &   0.4948(3) &   0.2541(4) &   0.078(3) \\
6  &             HMC &   8 &   4 &   2.0 &  0.5 &        0.85 &   1000 &   0.5297(5) &   0.2942(7) &   0.137(7) \\
7  &             HMC &   8 &   4 &   2.1 &  0.5 &        0.85 &  20200 &   0.5684(2) &   0.3439(3) &   0.327(4) \\
8  &             HMC &   8 &   4 &   2.2 &  0.5 &        0.84 &   1000 &   0.6041(5) &   0.3932(8) &   0.553(6) \\
9  &             HMC &   8 &   4 &   2.3 &  0.5 &        0.85 &   1000 &   0.6302(4) &   0.4296(6) &   0.679(4) \\
10 &             HMC &   8 &   4 &   2.4 &  0.5 &        0.85 &   1000 &   0.6507(4) &   0.4579(6) &   0.757(3) \\
\hline %
11 &            SLMC &   8 &   4 &   1.0 &  0.5 &        0.44 &    990 &   0.2559(3) &   0.0671(3) &   0.028(3) \\
12 &            SLMC &   8 &   4 &   1.2 &  0.5 &        0.47 &  41020 &  0.30368(6) &  0.09393(7) &  0.0319(5) \\
13 &            SLMC &   8 &   4 &   1.4 &  0.5 &        0.48 &  41930 &  0.35296(7) &  0.12676(7) &  0.0347(5) \\
14 &            SLMC &   8 &   4 &   1.6 &  0.5 &        0.46 &  41970 &  0.40492(5) &  0.16744(6) &  0.0424(5) \\
15 &            SLMC &   8 &   4 &   1.8 &  0.5 &        0.47 &  41570 &  0.46257(7) &   0.2205(8) &  0.0578(7) \\
16 &            SLMC &   8 &   4 &   1.9 &  0.5 &        0.47 &   5590 &   0.4949(3) &   0.2544(4) &   0.081(2) \\
17 &            SLMC &   8 &   4 &   2.0 &  0.5 &        0.46 &  41200 &  0.53001(7) &   0.2947(1) &    0.14(1) \\
18 &  SLMCup100 &   8 &   4 &   2.1 &  0.5 &        0.41 &  11920 &   0.5682(2) &   0.3437(3) &   0.324(3) \\
19 &            SLMC &   8 &   4 &   2.2 &  0.5 &        0.50 &  41180 &   0.6049(8) &   0.3944(1) &   0.559(1) \\
20 &            SLMC &   8 &   4 &   2.3 &  0.5 &        0.55 &  34670 &  0.63037(4) &  0.42958(6) &  0.6775(6) \\
21 &            SLMC &   8 &   4 &   2.4 &  0.5 &        0.60 &  34550 &  0.65066(3) &  0.45782(4) &  0.7558(5) \\
\hline %
\end{tabular}
\caption{
Same table with \tab\ref{tab:zero_temp} but for finite temperature with HMC and SLMC.
\label{tab:fin_temp}
}
\end{table}


\begin{figure}[h]
\begin{center}
\begin{minipage}{0.45\hsize}
\includegraphics[scale=0.45]{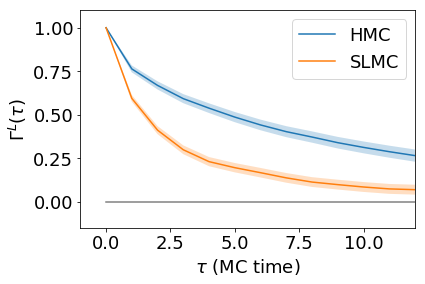}
\end{minipage}
\end{center}
  \caption{
  Autocorrelation function for the Polyakov loop
  for $\beta = 2.1$, $N_\sigma^3\times N_\tau = 8^3\times 4$, $\hat{m}= 0.5$.
  the number of the heatbath is 100 in SLMC.
  \label{fig:acfuncp_finite_SLMC_b21_m05_numbath100_output}}
\end{figure}

As we have seen, $\beta = 2.1$ is closes coupling to the critical (crossover) point.
\fig\ref{fig:acfuncp_finite_SLMC_b21_m05_numbath100_output}.
We choose the number of heatbath update as 100 in this case.

\section{Conclusion}
In this work, we develop self-learning Monte-Carlo algorithm for
lattice Yang-Mills theory with dynamical fermions in four dimensions.
We work with QC$_2$D with $n_f=4$ as an example
with zero and finite temperature.

We confirm that SLMC works for zero temperature runs even for the out of expansion radius of the hopping parameter expansion because of the exactness.
This is expected since SLMC itself is free from the choice of effective action.
The acceptance rate becomes low for small $\beta$ and large volume, but it can be fixed by adding more extended loops.
Our code for automatic generation of heatbath code will be published in another paper.

For finite temperature runs, we confirm that SLMC reproduces correct results with HMC, including higher-order moments of the Polyakov loop and the chiral condensate.
Our calculations indicate that QC$_2$D with $\hat{m}=0.5$ is in the crossover regime,  and we leave the precise determination of the order of phase transition to future study.

Our current algorithm uses heavy mass expanded effective action with the linear regression, which induces low efficiency of simulation for lighter mass.
This would be fixed by employing a neural network like \cite{Shen_2018}.

\section*{Acknowledgment}
Akio T would like to thank to
P. S. Bedaque 
W. Detmold, 
K. Orginos during a workshop ``A.I. FOR NUCLEAR PHYSICS WORKSHOP''
at Jefferson Lab
for notifying references and comments,
and S. Valgushev for fruitful discussion.
The work of Akio T was supported by the RIKEN Special Postdoctoral Researcher program
and partially supported by JSPS KAKENHI Grant Number JP20K14479.
The calculations were partially performed by the supercomputing systems SGI ICE X at the Japan Atomic Energy Agency. 
The numerical calculations were partially carried out on XC40 at YITP in Kyoto University.
YN was partially supported by JSPS- KAKENHI Grant Numbers 18K11345 and 18K03552. 
The work of Akinori T was partially supported by JSPS KAKENHI Grant Number 18K13548.

\bibstyle{apsrev4-1}
\bibliography{ref}

\end{document}